\documentclass[10pt,twocolumn,letterpaper]{article}

\usepackage[pagenumbers]{wacv}

\usepackage{graphicx}
\usepackage{amsmath}
\usepackage{amssymb}
\usepackage{float}
\usepackage{wrapfig}
\usepackage{picinpar}
\usepackage{cutwin}
\usepackage{caption}
\usepackage{subcaption}
\usepackage{url}
\usepackage{booktabs}
\usepackage{multirow}
\usepackage[absolute]{textpos}
\usepackage[pagebackref,breaklinks,colorlinks]{hyperref}
\usepackage[capitalize]{cleveref}
\Crefname{figure}{Figure}{Figures}
\Crefname{section}{Section}{Sections}
\Crefname{table}{Table}{Tables}
\newcommand{\mypara}[1]{\noindent\textbf{#1:}}
\newcommand{\question}[1]{\noindent\textbf{#1?}}
\newcommand{\targetmodel}{\mathcal{G}_\text{target}}
\newcommand{\attackmodel}{\mathcal{A}}
\newcommand{\auxiliarydataset}{\mathcal{D}_\text{aux}}
\newcommand{\trainingdataset}{\mathcal{D}_\text{train}}
\newcommand{\predictedstatus}{{y}_\text{query}}
\newcommand{\querysample}{\mathbf{x}_\text{query}}

\begin{document}

\begin{textblock}{16}(1.9,1)
To Appear in 2024 IEEE/CVF Winter Conference on Applications of Computer Vision (WACV), January 2024
\end{textblock}

\title{Generated Distributions Are All You Need for \\ Membership Inference Attacks Against Generative Models}

\author{
Minxing Zhang\textsuperscript{1}\ \ \
Ning Yu\textsuperscript{2}\ \ \
Rui Wen\textsuperscript{1}\ \ \
Michael Backes\textsuperscript{1}\ \ \
Yang Zhang\textsuperscript{1}
\\
\\
\textsuperscript{1}\textit{CISPA Helmholtz Center for Information Security} \ \ \ 
\textsuperscript{2}\textit{Salesforce Research}
}

\maketitle

%------------------------------------------------------------------------
\begin{abstract}
Generative models have demonstrated revolutionary success in various visual creation tasks, but in the meantime, they have been exposed to the threat of leaking private information of their training data.
Several membership inference attacks (MIAs) have been proposed to exhibit the privacy vulnerability of generative models by classifying a query image as a training dataset member or nonmember.
However, these attacks suffer from major limitations, such as requiring shadow models and white-box access, and either ignoring or only focusing on the unique property of diffusion models, which block their generalization to multiple generative models.
In contrast, we propose the first generalized membership inference attack against a variety of generative models such as generative adversarial networks, [variational] autoencoders, implicit functions, and the emerging diffusion models.
We leverage only generated distributions from target generators and auxiliary non-member datasets, therefore regarding target generators as black boxes and agnostic to their architectures or application scenarios.
Experiments validate that all the generative models are vulnerable to our attack.
For instance, our work achieves attack AUC $>0.99$ against DDPM, DDIM, and FastDPM trained on CIFAR-10 and CelebA.
And the attack against VQGAN, LDM (for the text-conditional generation), and LIIF achieves AUC $>0.90.$
As a result, we appeal to our community to be aware of such privacy leakage risks when designing and publishing generative models.\footnote{Code at \href{https://github.com/minxingzhang/MIAGM}{https://github.com/minxingzhang/MIAGM}}
\end{abstract}

%------------------------------------------------------------------------
\section{Introduction}
\label{section: introduction}

\begin{figure*}[!t]
\centering
\includegraphics[width=0.8\textwidth]{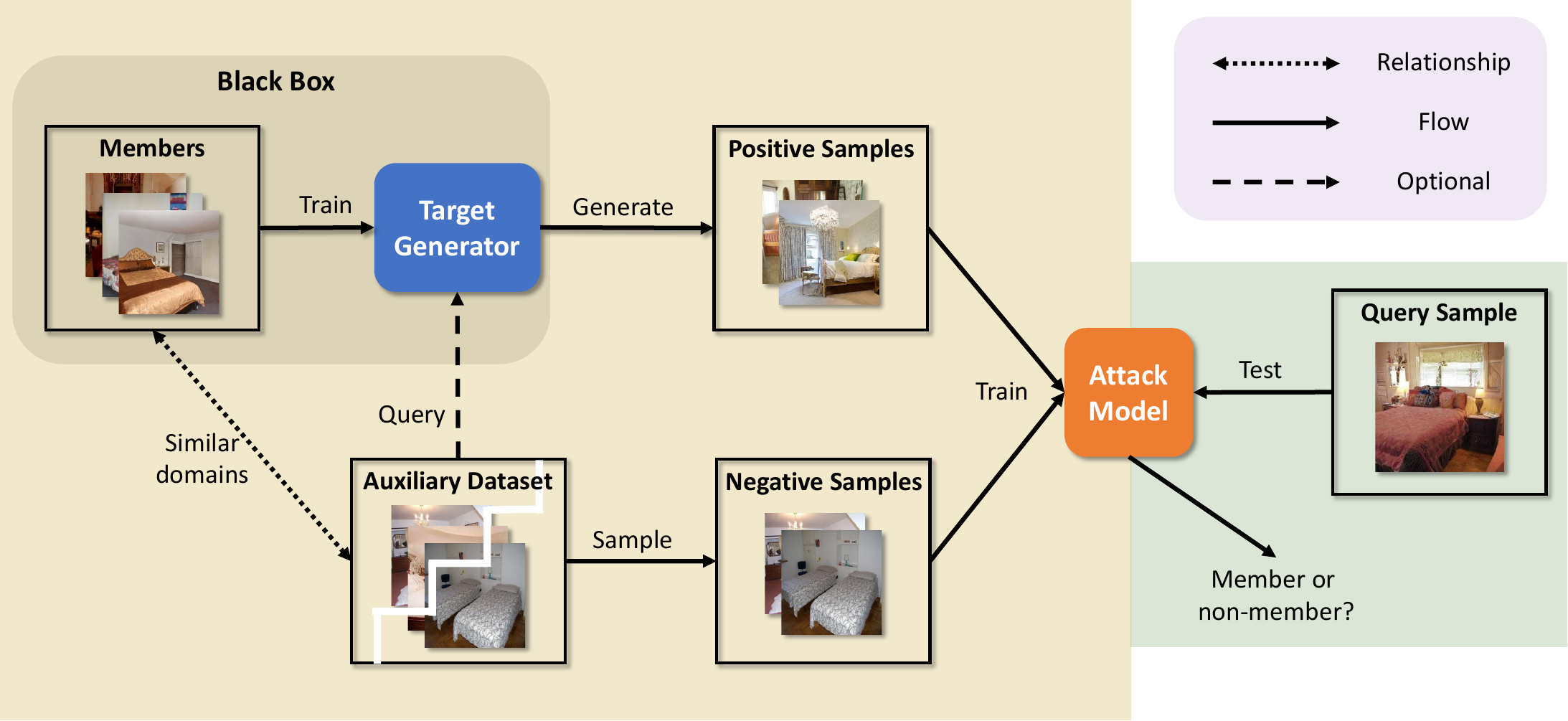}
\caption{The framework of our membership inference attack against a target generative model.}
\label{figure: attack pipeline}
\end{figure*}

The recent arms race in visual generation has reached a new peak.
Dall-E-2~\cite{RDNCC22}, Imagen~\cite{SCSLWDGAMLSHFN22}, Parti~\cite{YXKLBWVKYAHHPLZBW22}, and Stable Diffusion~\cite{stablediffusion}, driven by big data and empowered by deep generative models, have emerged one after another in a short period of time to showcase their improved power.
In the backend, generative adversarial networks (GANs)~\cite{gan}, [variational] autoencoders ([V]AEs)~\cite{ae,KW142}, implicit functions (IFs)~\cite{nerf}, and the emerging denoising diffusion probabilistic models (DDPMs)~\cite{ddpm} are four representative milestones.
While we enjoy the benefits of these technologies, the threats of privacy leaks~\cite{CYZF20,AGA21,HP21,ZCSZ22} simultaneously come with them and can be easily overlooked.
And privacy leakages will get severe especially when data are misused, e.g. the DeepMind project.\footnote{\href{https://news.sky.com/story/google-received-1-6-million-nhs-patients-data-on-an-inappropriate-legal-basis-10879142}{https://news.sky.com/story/google-received-1-6-million-nhs-patients-data-on-an-inappropriate-legal-basis-10879142}}

To this end, we aim to expose the potential privacy risks from generative models through the lens of membership inference attacks (MIAs), which is by far one of the most popular and powerful techniques~\cite{SSSS17,SZHBFB19}.
MIAs target to infer whether a query sample is in the training set of the target model or not.
Most existing MIAs aim at discriminative models~\cite{CTCP21,ZRWRCHZ21,LZ21,CCNSTT22}, which cannot directly extend to generative models due to the distinct architectures.
A few existing MIAs explore privacy risks in generative models~\cite{HMDC19,CYZF20,AGA21,SecMI}.
This kind of attack is worth the attention due to multifold reasons:
Malicious adversaries can leverage MIAs to cause severe consequences.
For instance, GANs are applicable to medical images~\cite{YWB19} where the training data (i.e., the medical history of patients) is sensitive, and the privacy leakage may threaten the target individual's life.
Meanwhile, MIAs can also be used for benign purposes such as auditing, i.e. quantitatively assessing whether licensed data samples are illegally pirated for generator training or whether private information is misused without permission.
For instance, everyone can audit whether their private photo is involved in any dataset on which generative models can be trained.\footnote{\href{https://twitter.com/LapineDeLaTerre/status/1570889343845404672}{https://twitter.com/LapineDeLaTerre/status/1570889343845404672}}
Moreover, MIAs are commonly used as the building block for more sophisticated attacks, which makes it an important topic and has attracted lots of attention in recent years.
For instance, MIAs can be an important module for the holistic risk assessment of generative models~\cite{LWHSZBCFZ22}.

However, the applicable scope of existing MIAs against generative models is limited in two aspects.
First, these works are designed under certain attack assumptions, e.g. requiring white-box access and shadow models~\cite{LF20,AGA21}.
Second, they fail to fit a variety of generative applications~\cite{CYZF20,SecMI}.
Concretely, most of these works ignore the diversity property of DDPMs, which hinders their application in state-of-the-art generative models;
on the contrary, a few recent works only take advantage of the unique features of DDPMS, but do not take other generative models into consideration, i.e., GANs, [V]AEs, and IFs.
These limitations motivate researchers to propose more practical and generalized MIAs.

In this paper, utilizing the generated distribution learned from the target generator's outputs, we propose a generalized membership inference attack against various generative models in a variety of applications.
Compared to existing attacks, our method has three main advantages:
(i) \textbf{Relaxed assumptions}. 
Our work assumes that the attacker only has black-box access to target models.
That is, the attacker only needs to query target models in an API manner.
(ii) \textbf{Computationally efficient}. 
Contrary to existing attacks, our method does not require training shadow models, which makes the method computationally efficient.
(iii) \textbf{Generalizability}.
Our attack can be applied to a variety of generative models including but not limited to the state-of-the-art DDPMs, whereas previous MIA designs are specified to certain model variants, for example, Azadmanesh et al. only evaluate privacy risks of GANs~\cite{AGA21} and Duan et al. only focus on DDPMs~\cite{SecMI}.
Besides, generalizability also benefits from our relaxed assumptions.
The black-box access represents a common scenario in practice, based on which our attack is independent of generator architectures, indicating that our attack is applicable to various generative models.

As \cref{figure: attack pipeline} shows, we use generated samples as training positives and auxiliary samples as training negatives.
In particular, when attacking conditional generators, the inputs to the target generators are also sampled from the auxiliary dataset but disjoint with training negatives.
In this case, still, the generated samples leak the information of members that were used to train the target generator.
This relies on the target generator memorizing the training distribution due to overfitting.
Then the distribution of generated images is to some extent an approximation of the training distribution.
Thus, the attack model can indirectly learn the target training distribution from the generated images, and further predict the membership of the target generator.

Regarding auxiliary datasets, they are involved due to a currently prevalent setting.
That is, to train large-scale generative models, model owners usually collect data as much as possible, and use the whole dataset in the training process.
In this case, the adversary's accessible data are all target training data, i.e., members, and no left and unused data from the same dataset can be used as negative training samples for the attack model.
To this end, the adversary collects auxiliary datasets of similar domains as member samples.

We extensively evaluate our MIA method against four widely utilized generative model frameworks (GANs, [V]AEs, IFs, DDPMs) covering ten generation applications (unconditional~\cite{ddpm,ddim,fastdpm}, class-conditional~\cite{taming}, text-conditional~\cite{RBLEO22} and semantic-conditional~\cite{cc-fpse} generation, image colorization~\cite{colorization}, super resolution~\cite{liif}, image inpainting~\cite{mat,RBLEO22}, stylization~\cite{swappingAuroencoder}, denoising~\cite{mprnet}, and artifact reduction~\cite{mprnet}).
For the pioneering attacks against unconditional DDPM, we cover the original DDPM~\cite{ddpm}, DDIM~\cite{ddim}, and FastDPM~\cite{fastdpm}.
Experimental results show the efficacy of our attacks on all the scenarios.
For instance, our work achieves attack AUC $>0.99$ against DDPM, DDIM, and FastDPM trained on CIFAR-10 and CelebA.
And the attack against VQGAN, LDM (in the application of text-conditional generation), and LIIF achieves AUC $>0.90.$
Further studies demonstrate that the generative models are still vulnerable to our attack even with a limited query budget, and the transferability makes our attack a real threat in real-world scenarios.

To better understand our work, we further discuss the membership boundary, the overlap between auxiliary and member samples, and the feasibility of generating training negatives.

\mypara{Our contributions} \\
\noindent -
We propose the first generalized MIA against various generative models (including GANs, [V]AEs, IFs, and state-of-the-art DDPMs) in various visual generation applications. \\
\noindent -
Our work is the first to exploit generated distributions learned from the target generator's outputs for membership inference.
Generated distributions are the common property of generators and are independent of generator architectures.
Meanwhile, our work requires fewer assumptions and is more computationally efficient than previous attacks.
These advantages earn our attack a practical and general application scope. \\
\noindent -
We empirically demonstrate the efficacy of our attack in all the scenarios, which in turn validate the generalizability and practical usage of our attacks. 
Besides, further studies showcase more advantageous properties, such as effectiveness with a limited query budget and transferability, which enhance the practical significance of the attack. \\
\noindent -
Our work fills the blank of understanding the common privacy risks of various generative models, which motivates researchers to take privacy threats into concern when designing and publishing generative models.

%------------------------------------------------------------------------
\section{Related Works}
\label{section: related works}

\mypara{Generative models and applications}
In this work, we consider the widely used generative models, i.e.,
generative adversarial networks (GANs)~\cite{ACB17,WLZTKC18,ZGMO18,BDS19,PWSCL21} which utilize the arms race between generators and discriminators to improve the quality of generated images;
[variational] autoencoders ([V]AEs)~\cite{KW142,beta-vae,SLY15,OVK17,TBGS18,CLGD18} which restore the input images from latent representations sampled from estimated normal distributions;
implicit functions (IFs)~\cite{nerf,liif,MolGan,ml16} which circumvent the need for an explicit likelihood;
and diffusion probabilistic models (DDPMs)~\cite{ddpm,ddim,fastdpm,RBLEO22,RDNCC22,SCSLWDGAMLSHFN22} which denoise noise images step by step until being clean images.
These above models cast significant success towards a variety of visual generation applications, including unconditional~\cite{ddpm,ddim,fastdpm}, class-conditional~\cite{taming}, text-conditional~\cite{RBLEO22} and semantic-conditional~\cite{cc-fpse} generation, image colorization~\cite{colorization}, super resolution~\cite{liif}, image inpainting~\cite{mat,RBLEO22}, stylization~\cite{swappingAuroencoder}, denoising~\cite{mprnet}, and artifact reduction~\cite{mprnet}.
And these applications require different kinds of inputs, i.e., noises, texts, and images.
We, therefore, nest our experiments in these applications to investigate the efficacy and generalizability of our attacks against various generative models.

\mypara{Membership inference attacks against generative models}
Many existing MIAs against generative models focus on GAN- and [V]AE-based generative models, which limits the applicable scope of membership inference.
For instance, Hayes et al. aim at several GANs and a single [V]AE~\cite{HMDC19};
Hilprecht et al. propose two types of MIAs, one applicable to GANs and [V]AEs while the other specific for [V]AEs~\cite{HHB19};
Chen et al. present the first taxonomy of membership inference, which mainly focuses on GANs~\cite{CYZF20};
and Azadmanesh et al. conduct a white-box membership inference against GANs~\cite{AGA21}.
Unfortunately, these works cannot generalize well to all aforementioned generative models and applications.
For instance, due to the diverse property of DDPMs, the attacks proposed by Chen et al. cannot achieve performances as strong as against GANs.
Recently, a few works explore the privacy issues of DDPMs.
For instance, Duan et al. utilize step-wise errors to infer membership~\cite{SecMI}.
However, this work relies on the unique feature of DDPMs, which is not available to other generative models, i.e., GANs, [V]AEs, and IFs;
Carlini et al. present the training data extraction of DDPMs~\cite{CHNJSTBIW23}, which is another popular topic regarding privacy leakage but aims at a different goal of our attack.
To propose a generalized membership inference attack, we take advantage of generated distributions that are architecture-agnostic.
Thus our work is applicable to various generative models and applications.

%------------------------------------------------------------------------
\section{Our Attack}
\label{section: our attack}

\subsection{Problem Statement}
\label{subsection: problem statement}

In this paper, we study the problem of membership inference attacks against generative models.
The goal is to infer whether a query image $\querysample$ belongs to the training set (i.e., members) of the target generative model $\targetmodel$.
The attack $\mathcal{A}$ can be formulated as follow:
\begin{equation}
\label{equation:l1_loss}
    \mathcal{A}(\querysample|
    \targetmodel,\mathcal{K})\rightarrow\{\text{member},\text{non-member}\}
\end{equation}
where $\mathcal{K}$ denotes extra information known to the attacker, and members and non-members are the data samples used and not used to train the target generator, respectively.
In our work, we assume the attacker has access to an auxiliary dataset $\auxiliarydataset$ of similar domains as the training dataset $\trainingdataset$ (i.e., members), which will be explained in \cref{subsection: auxiliary datasets}.
Hereby $\mathcal{K}\doteq\auxiliarydataset.$
The auxiliary dataset $\auxiliarydataset$ is separated into two disjoint parts, i.e., $\mathcal{D}_\text{aux}^\text{in}$ for querying the conditional target generator and $\mathcal{D}_\text{aux}^\text{out}$ as training negatives for the attack model, which will be detailed in the following.

\subsection{Methodology}
\label{subsection: methodology}

\cref{figure: attack pipeline} depicts the general framework of our attack.
The attack process consists of two steps: image generation and membership inference, with the former one being the core step of our attack.

\mypara{Image generation}
To infer the membership of the target generator, the attacker needs to collect training positives and negatives for the establishment of their attack model.
As mentioned in \cref{subsection: problem statement}, the training negatives are derived from $\mathcal{D}_\text{aux}^\text{out}$, formally:
\begin{equation}
    \label{eq:nonmem}
    \forall x\in\mathcal{D}_\text{aux}^\text{out},\quad \text{we have} \quad (x,\text{non-member})
\end{equation}
Thus, the current challenge is how to obtain quantities of training positives, since the member samples are inaccessible to the adversary.
Due to overfitting, the target generator memorizes its training distribution.
In that case, the generated images can reflect the pattern of real members to a large extent.
Consequently, the samples generated by the target generator are used as the training positives:
(i)
For unconditional generation tasks, we query the generative model with Gaussian noise, i.e., $z \sim \mathcal{N}(0,1)$.
Then we obtain a positive training pair, i.e., $(\targetmodel(z),\text{member})$, for the attack model;
(ii)
For conditional generation tasks, the target generator receives information to guide the generation process.
As an example, we illustrate the case where the input is image.
As mentioned in \cref{subsection: problem statement}, we query the target generator using the data sampled from $\mathcal{D}_\text{aux}^\text{in}$.
Then the training positives are formulated as:
\begin{equation}
    \label{eq:member}
    \forall x\in\mathcal{D}_\text{aux}^\text{in},\quad \text{we have} \quad (\targetmodel(x),\text{member})
\end{equation}
Note that our attack is architecture- and task-agnostic, thus the input to the generative model could be other types of information, like text sentences, images with artifacts, and grayscale images.

\mypara{Membership inference}
The adversary establishes a binary classifier as the attack model $\attackmodel$ in a supervised way using the training positives and negatives constructed in the previous step.
Then, the adversary can infer the membership status of a query sample $\querysample$.
Note that, in the inference procedure, the adversary does not need to interact with the target generative model.
Specifically, the adversary directly queries the trained attack model $\attackmodel$ with the query sample $\querysample$.
And the membership status is inferred according to the output of the attack model.
This process can be formulated as:
\begin{equation}
    \label{equation:inference}
\attackmodel(\querysample) \rightarrow \predictedstatus
\end{equation}
where $\predictedstatus\in\{0,1\}$ indicates the membership status of the query sample $\querysample$.

Different from previous works that rely on strong assumptions (i.e., white-box access to the target generator), we only require black-box access, which means that interactions with the generator by the attacker can only happen through an API manner.
And thanks to the black-box access, our attack is architecture-agnostic and can be generalized to a variety of generative models.
To better understand our work, we empirically validate this relaxed and practical assumption by extensive experiments in \cref{section: evaluation}, and clarify the superiority of our attack as follows:

\mypara{The effectiveness}
Our attack utilizes generated distributions learned from the target generator's outputs to conduct membership inference.
This is because overfitting helps the target generator remember the distribution of its training data.
Thus the distribution of generated images can be regarded as an approximate of the training distribution.
In that case, the attack model can indirectly learn the training distribution from the generated images, and further predict the membership of the target generator.
Note that, image generation is a basic function of generative models, so generated images are of course available via black-box access in all generative applications.

\mypara{The advantages}
(i) \emph{Relaxed assumptions.}
Our work only assumes black-box access to the target generator, which is a prevalent setting in practice and thus largely enhances the attacker's ability compared to white-box access.
(ii) \emph{Computationally efficient.}
Different from many existing works, no requirement of training shadow models makes our approach more computationally efficient.
(iii) \emph{Generalizability.}
Not like previous MIAs that ignore or only focus on DDPMs, our method is applicable to all the widely used generative models, i.e., GANs, [V]AEs, IFs, and DDPMs.

\begin{table*}[!t]
    \caption{The settings of generative applications, target generative models, and member and auxiliary datasets.}
    \resizebox{\textwidth}{!}{
    \centering
    \begin{tabular}{l | c c c | c c | c c }
        \toprule
        \multirow{3}{*}{\textbf{Application}} & \multicolumn{3}{c|}{\multirow{2}{*}{\textbf{Target generative model}}} & \multicolumn{4}{c}{\textbf{Attack binary classifier}} \\
         & & & & \multicolumn{2}{c|}{\textbf{Training images}} & \multicolumn{2}{c}{\textbf{Testing images}} \\
        & \textbf{Technique} & \textbf{Framework} & \textbf{Training/Member dataset} & \textbf{Positive} & \textbf{Negative} & \textbf{Positive} & \textbf{Negative} \\
        \midrule
        \multirow{4}{*}{Unconditional} & \multirow{4}{*}{DDPM~\cite{ddpm} DDIM~\cite{ddim} FastDPM~\cite{fastdpm}} & \multirow{4}{*}{DDPM} & CIFAR-10 & \multirow{4}{*}{$\mathcal{G}$(noise)} & STL-10 & & \\
         &  & & CelebA & & UTKFace & & \\
         &  & & LSUN-Bedroom & & Wild Bedroom & & The \\
         &  & & LSUN-Church & & Wild Church & The & same \\
        \cmidrule{1-6}
        Class-conditional & VQGAN~\cite{taming} & GAN & ImageNet & $\mathcal{G}$(class) & Open Images & same & as \\
        Text-conditional & LDM~\cite{RBLEO22} & DDPM & LAION & $\mathcal{G}$(COCO text) & COCO & as & attacker \\
        Semantic-conditional & CC-FPSE~\cite{cc-fpse} & GAN & COCO & $\mathcal{G}$(ADE20K semantics) & ADE20K & target & training \\
        Image colorization & Colorization~\cite{colorization} & [V]AE & ImageNet & $\mathcal{G}$(Open Images gray) & Open Images & generator & negative \\
        Super resolution & LIIF~\cite{liif} & IF & CelebA-HQ & $\mathcal{G}$(UTKFace low res) & UTKFace & training & but \\
        \multirow{2}{*}{Image inpainting} & MAT~\cite{mat} & GAN+[V]AE & CelebA-HQ & $\mathcal{G}$(UTKFace incomplete) & UTKFace & dataset & disjoint \\
        & LDM~\cite{RBLEO22} & DDPM & LAION & $\mathcal{G}$(COCO text) & COCO & & split \\
        Stylization & SwappingAutoencoder~\cite{swappingAuroencoder} & GAN+[V]AE & LSUN-Church & $\mathcal{G}$(Wild Church) & Wild Church &  & \\
        Denoising & MPRNet~\cite{mprnet} & [V]AE & SIDD & $\mathcal{G}$(ImageNet noisy) & ImageNet &  & \\
        Artifact reduction & MPRNet~\cite{mprnet} & [V]AE & SIDD & $\mathcal{G}$(ImageNet artifact) & ImageNet & & \\
        \bottomrule
    \end{tabular}}
    \label{table: external scenarios}
\end{table*}

\subsection{Auxiliary Datasets}
\label{subsection: auxiliary datasets}

As aforementioned, our approach involves an auxiliary dataset of similar domains as the training dataset of the target generator. We will specify this in this section.

\question{Why use auxiliary datasets}
A currently prevalent setting is that the owners of large-scale generative models usually collect data as much as possible, and use the whole dataset for training.
In that case, the adversary's accessible data are all target training data, i.e., members.
That is, no left and unused data from the same dataset can be used as training negatives for the attack model.
To this end, the adversary samples training negatives from auxiliary datasets of similar domains as member samples.

\question{Why similar domains}
The decision of similar domains is because the adversary is hard to obtain an auxiliary dataset of the exact domains of the target generator, and the discussion of obviously different domains is trivial.
Specifically, from the generated images, the adversary can easily estimate the target domains, e.g., generated ImageNet~\cite{TinyImageNet} images depict dogs, cats, etc.
However, the adversary's ability might be limited by query times/cost, thus the estimated domains would only be similar but not exactly the same.
And even if possible, it is hard to guarantee the existence of available datasets of the exact domains, e.g. LAION.
Thus, aiming at similar domains instead of the exact ones brings a broader application scope.
On the other hand, according to the definition, members are the training samples of the target generator, which means all samples that are not used for training are non-members.
In that case, the images of obviously different domains will be easily classified as non-members, so the challenge is about similar domains.

\question{How to construct}
Based on the estimated domain, there are two ways to construct an auxiliary dataset:
(i)
Different datasets have domain overlaps, e.g., CIFAR-10~\cite{CIFAR} and STL-10~\cite{STL10} both contain the airplane category.
Thus, the adversary can utilize an existing dataset of similar domains.
Note that, the larger the domains of the auxiliary dataset, the more the possibility of covering the target domains, and of course the better.
For instance, even though CIFAR-10 and Open Images~\cite{openimages} both describe real-world objects, we use Open Images as the auxiliary dataset for ImageNet since it contains more categories;
(ii)
The alternative is to manually collect auxiliary samples when no existing dataset of similar domains is available.
For instance, the bedroom images collected from the internet are used as the auxiliary dataset for LSUN-Bedroom~\cite{lsun}.

Ideally, there is no overlap between the auxiliary dataset and the generator's training data (i.e., members), otherwise, the (part of) training negatives sampled from the auxiliary dataset should be labeled as ``members'', which is contrary to the previous setting.
However, since the member samples are inaccessible, the adversary cannot check and guarantee non-overlap.
To this end, we conduct more explorations in \cref{section: discussion} to understand the influence of overlap.

%------------------------------------------------------------------------
\section{Evaluation}
\label{section: evaluation}

\begin{figure}[!t]
\centering
\subfloat[(\textbf{CIFAR-10}, STL-10)]{
    \includegraphics[width=0.45\textwidth]{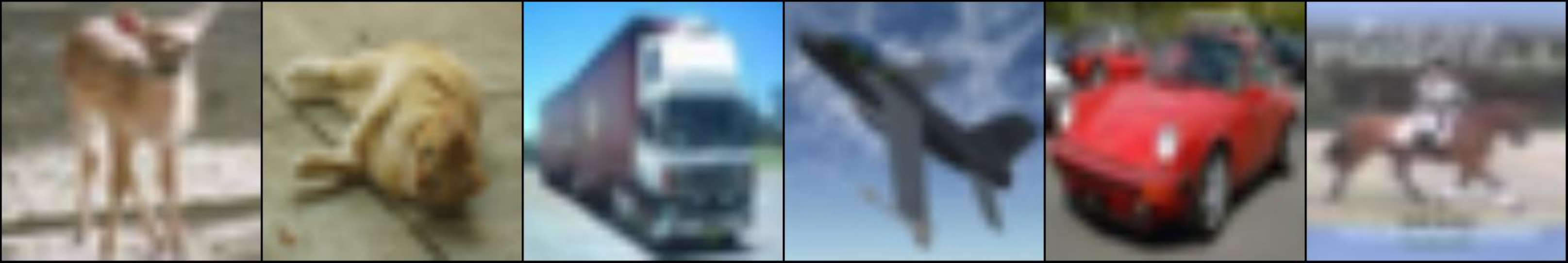}
}

\subfloat[(CIFAR-10, \textbf{STL-10)}]{
    \includegraphics[width=0.45\textwidth]{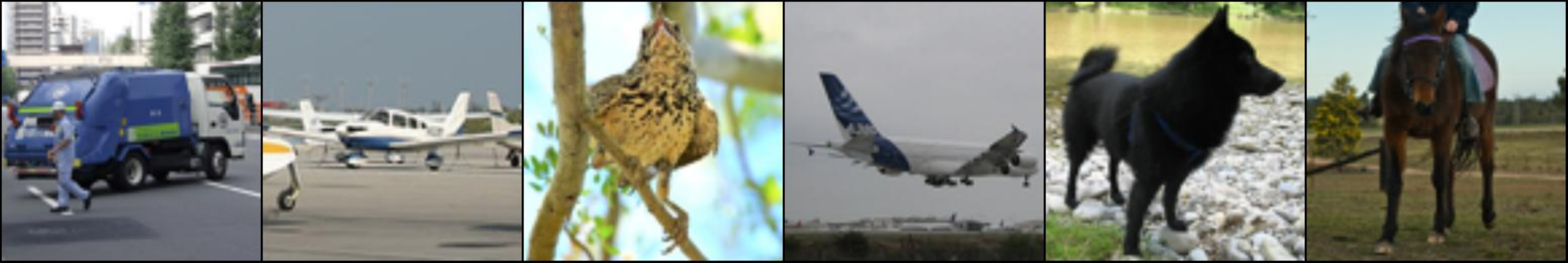} 
}

\subfloat[(\textbf{CelebA}, UTKFace)]{
    \includegraphics[width=0.45\textwidth]{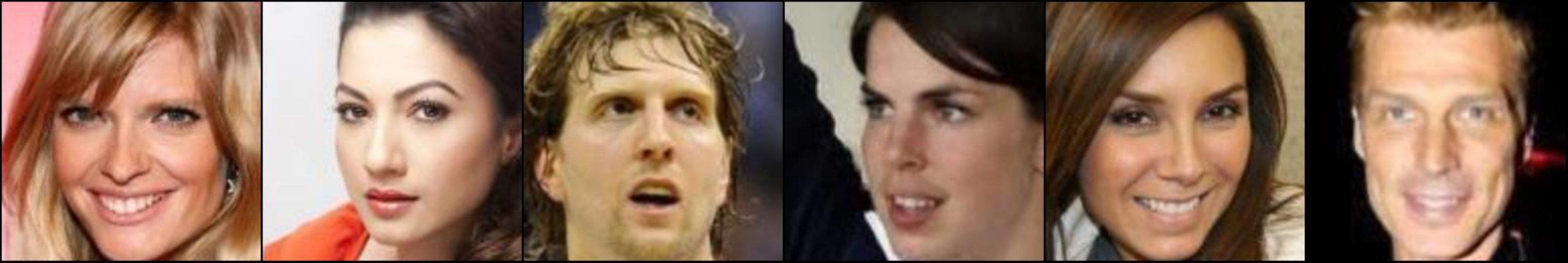} 
}

\subfloat[(CelebA, \textbf{UTKFace})]{
    \includegraphics[width=0.45\textwidth]{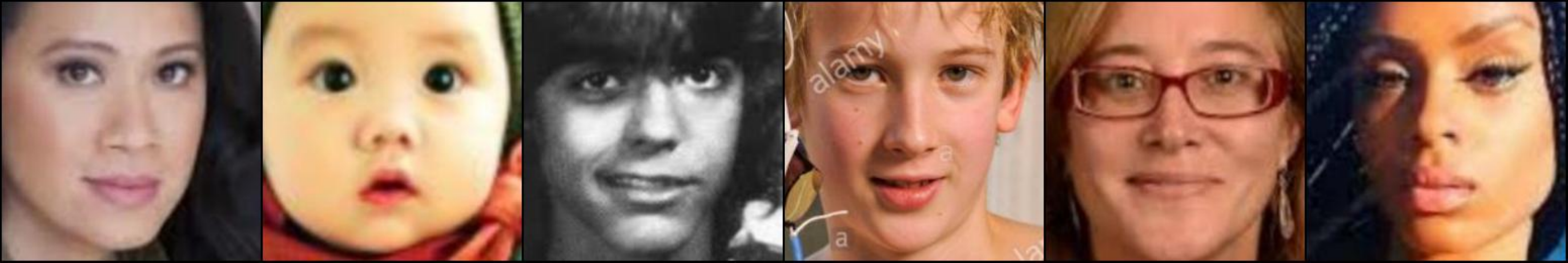}
}

\subfloat[(\textbf{LSUN-Bed}, Wild Bed)]{
    \includegraphics[width=0.45\textwidth]{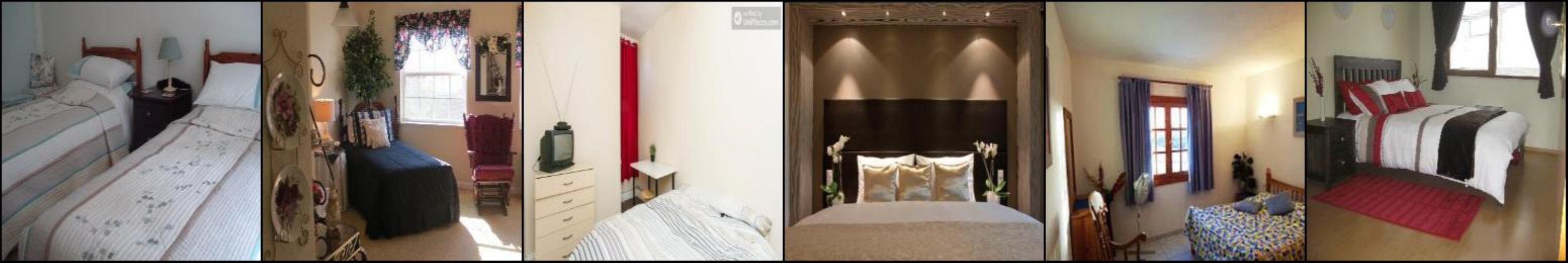}
}

\subfloat[(LSUN-Bed, \textbf{Wild Bed})]{
    \includegraphics[width=0.45\textwidth]{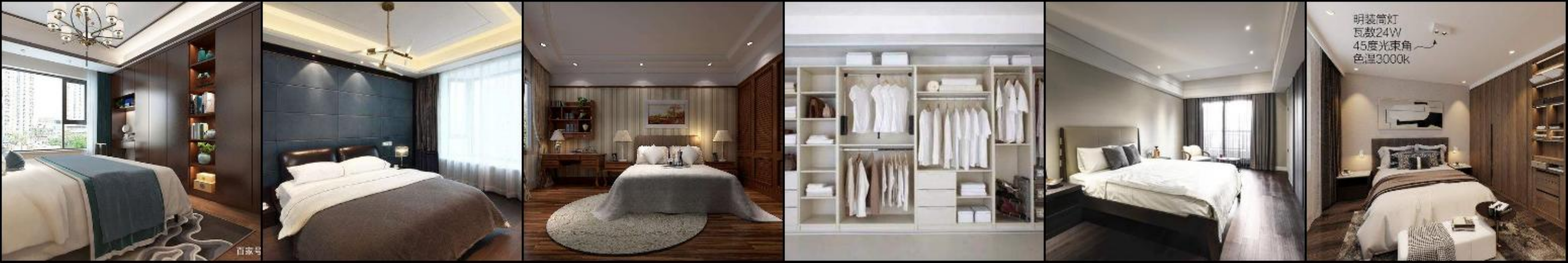}
}

\subfloat[(\textbf{LSUN-Church}, Wild Church)]{
    \includegraphics[width=0.45\textwidth]{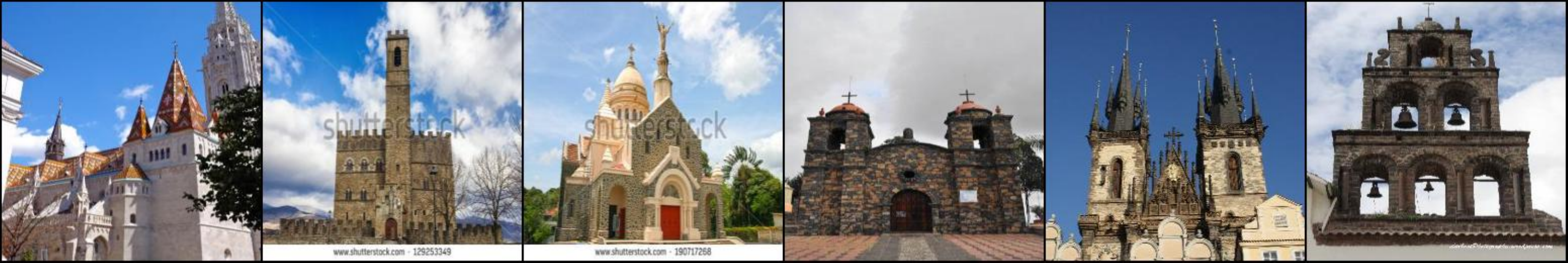}
}

\subfloat[(LSUN-Church, \textbf{Wild Church})]{
    \includegraphics[width=0.45\textwidth]{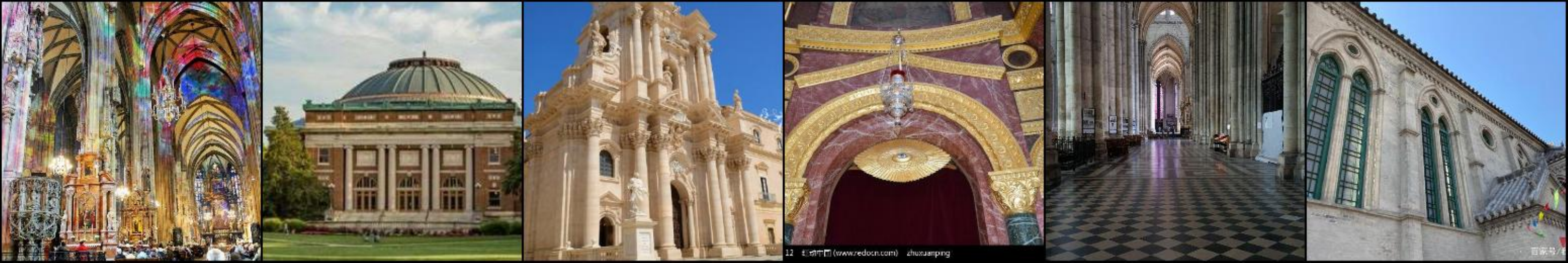}
}
\caption{The image examples of different (member, auxiliary) dataset pairs, where the images are sampled from the \textbf{boldface dataset}.}
\label{figure: dataset examples} 
\end{figure}

\subsection{Evaluation Settings}
\label{subsection: evaluation settings}

\mypara{Generative applications and models}
To comprehensively understand our work, we conduct the evaluation on ten generative applications, i.e., unconditional, class-conditional, text-conditional and semantic-conditional generation, image colorization, super resolution, image inpainting, stylization, denoising, and artifact reduction, as listed in the first column of \cref{table: external scenarios}.
Extensive generative techniques of these applications are involved to measure the attack performance in various scenarios, as shown in the second column of the table.
Specifically, DDPM~\cite{ddpm}, DDIM~\cite{ddim} and FastDPM~\cite{fastdpm} for unconditional generation;
VQGAN~\cite{taming} for class-conditional generation;
LDM~\cite{RBLEO22} for text-conditional generation;
CC-FPSE~\cite{cc-fpse} for semantic-conditional generation;
Colorization~\cite{colorization} for image colorization;
LIIF~\cite{liif} for super resolution;
MAT~\cite{mat} and LDM for image inpainting;
SwappingAutoencoder~\cite{swappingAuroencoder} for stylization;
MPRNet~\cite{mprnet} for both denoising and artifact reduction.
Regarding the backbone frameworks of these techniques, four widely used generative models are covered, which corresponds to the third column of the table, i.e., GANs (including VQQGAN, CC-FPSE, MAT, and SwappingAutoencoder), [V]AEs (including Colorization, MAT, SwappingAutoencoder, and MPRNet), IFs (including LIIF), and DDPMs (including DDPM, DDIM, FastDPM, and LDM).

\mypara{Member and auxiliary datasets}
As mentioned in \cref{section: our attack}, the adversary accesses an auxiliary dataset for each training/member dataset of target generators.
Thus, we introduce the dataset pairs of $($member, auxiliary$)$ used for different applications in the other columns of \cref{table: external scenarios}, and show some examples in \cref{figure: dataset examples}:
$($CIFAR-10~\cite{CIFAR}, STL-10~\cite{CNL11}$)$ $\rightarrow$ unconditional generation;
$($CelebA(-HQ)~\cite{LLWT15}, UTKFace~\cite{ZSQ17}$)$ $\rightarrow$ unconditional generation, super resolution and image inpainting;
$($LSUN-Bedroom~\cite{lsun}, Wild Bedroom$)$ $\rightarrow$ unconditional generation;
$($LSUN-Church, Wild Church$)$ $\rightarrow$ unconditional generation and stylization;
$($ImageNet\cite{DDSLLF09}, Open Images~\cite{openimages}$)$ $\rightarrow$ class-conditional generation and image colorization;
$($LAION~\cite{SVBKMKCJK21}, COCO~\cite{coco}$)$ $\rightarrow$ text-conditional generation and image inpainting;
$($COCO, ADE20K~\cite{ade20k}$)$ $\rightarrow$ semantic-conditional generation;
and $($SIDD~\cite{sidd}, ImageNet$)$ $\rightarrow$ denoising and artifact reduction.
These datasets are further introduced in \cref{appendix: datasets}.
Even in some cases, the domains of auxiliary datasets are not exactly the same as the members' as explained in \cref{subsection: auxiliary datasets}, e.g., frog images in CIFAR-10 but not in STL-10, surprisingly, our attack still works well in all applications as shown in \cref{subsection: results}.

\mypara{Attack model}
The attack model is established based on a pre-trained ResNet18~\cite{HZRS16}, of which the last linear layers are fitted for binary classification, i.e., inferring the query sample belonging to members or non-members.
More training details are shown in \cref{appendix: training details}.

\mypara{Evaluation metric}
We utilize AUC as the metric to evaluate attacks~\cite{SZHBFB19,CYZF20,LZ21,ZRWRCHZ21}, of which the value is in the range of $[0.0, 1.0]$, and the higher the more effective.
Specifically, during the testing, only the images used to train the target generator are regarded as positive samples while the others are all regarded as negative samples.

\subsection{Results}
\label{subsection: results}

\begin{figure*}[!t]
\centering
\subfloat[Unconditional generation tasks.]{
    \centering
    \label{subfigure: unconditional results}
    \includegraphics[width=0.45\textwidth]{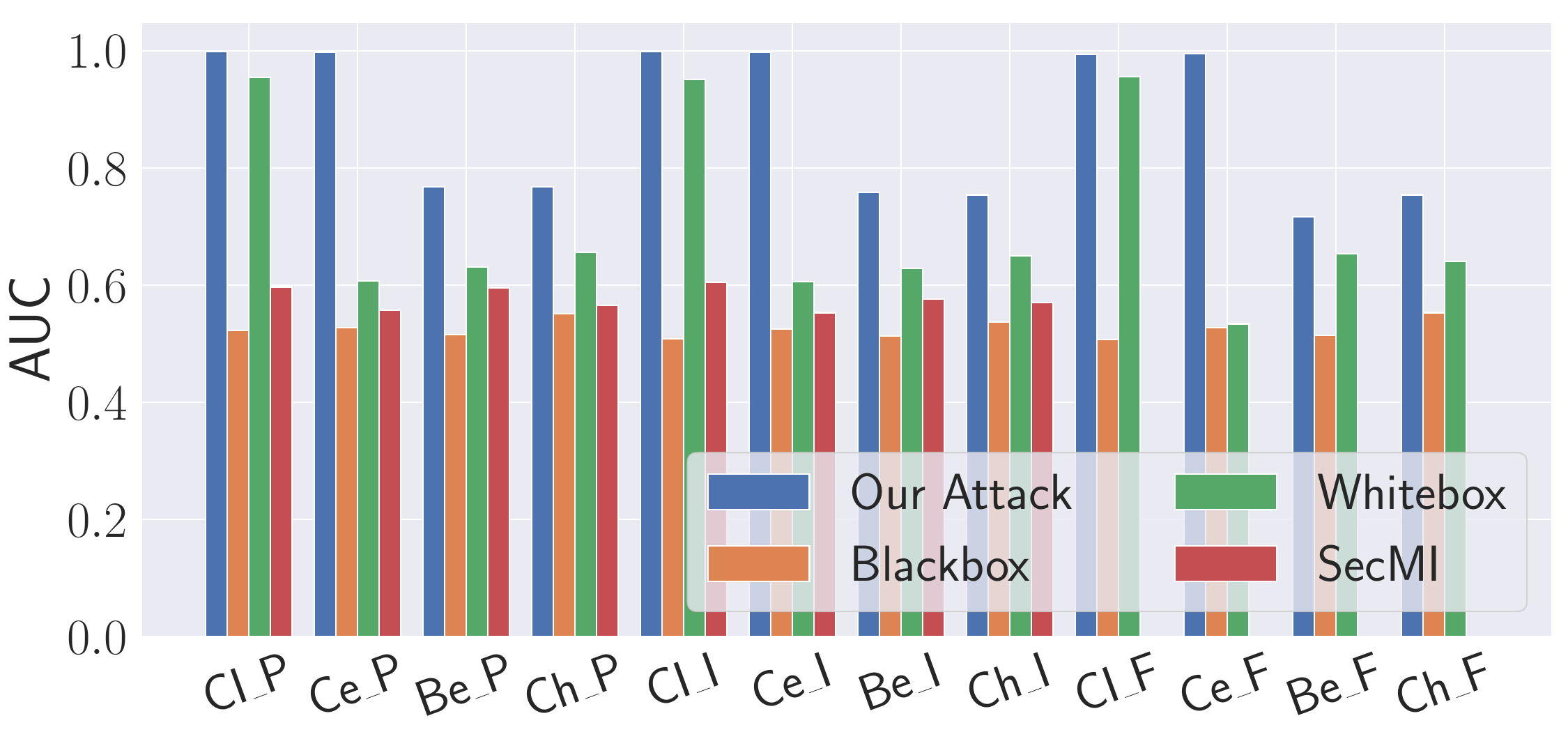}
}
\subfloat[Conditional generation tasks.]{
    \centering
    \label{subfigure: conditional results}
    \includegraphics[width=0.45\textwidth]{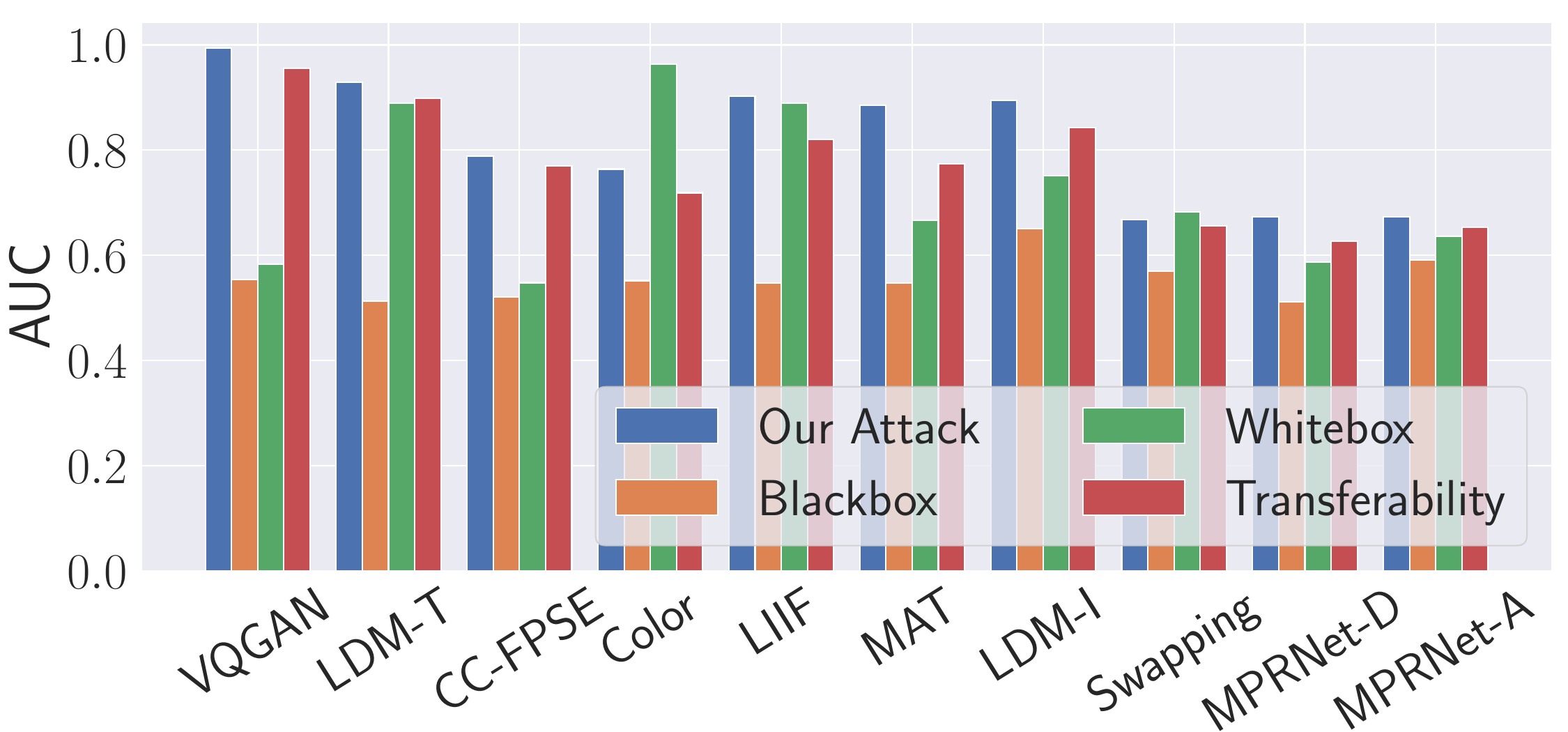}
}
\caption{The attack performances in various generative applications.
In \cref{subfigure: unconditional results}, we use CI/Ce/Be/Ch\_P/I/F notation in the x-axis to denote the generator DDPM/DDIM/FastDPM trained on the dataset CIFAR-10/CelebA/Bedroom/Church.
In the x-axis of \cref{subfigure: conditional results}, ``Color'' and ``Swapping'' are short for ``Colorization'' and ``SwappingAutoencoder'', ``LDM-T'' and ``LDM-I'' represent LDM used for text-conditional generation and image inpainting, and ``MPRNet-D'' and ``MPRNet-A'' are MPRNet used for denoising and artifact reduction.}
\label{figure: effective on generative applications}
\end{figure*}

In this part, we evaluate the effectiveness of our attack on ten generative applications and depict the experimental results in \cref{figure: effective on generative applications}.
To better show the superiority, we compare our attack with the black- and white-box attacks of Chen et al.~\cite{CYZF20} which are one of the most popular MIAs against generative models.
Moreover, we compare our attack against DDPM and DDIM with SecMI~\cite{SecMI} which is a recently proposed MIA work specific to diffusion models.

\cref{figure: effective on generative applications} shows that all generative scenarios are vulnerable to our attack.
For instance, our attack depicts consistent efficacy with AUC $>0.99$ on DDPM, DDIM, and FastDPM trained on CIFAR-10 or CelebA as shown in \cref{subfigure: unconditional results}.
And the attack achieves AUC $>0.9$ on VQGAN, LDM-T, and LIIF trained on ImageNet, LAION, and CelebA-HQ, respectively, as shown in \cref{subfigure: conditional results}.
Meanwhile, the black-box baseline~\cite{CYZF20} only achieves AUC $<0.6.$
The obvious advantages over the baselines well demonstrate the effectiveness of our work in various applications, which further indicates that our attack can effectively expose the private information of various generative models.

An interesting finding appears in conditional generation tasks.
Specifically, our generated images (i.e., training positives for the attack model) are derived by querying the target generator using a split of an auxiliary dataset, which indicates the generated images are associated with auxiliary data.
Even in such a non-trivial setting, surprisingly, the boundary learned by the attack model from the generated and auxiliary samples can also serve as an effective boundary during the testing to separate member samples and another split of auxiliary samples.
For instance, our attack on LIIF (for the super resolution) achieves an AUC of 0.902.
This shows the advantage of utilizing generated distributions to learn the information of member samples as mentioned in \cref{subsection: methodology}.
In other words, the generated images, even if not only associated with member samples, can still expose the membership status of member samples, which introduces a large applicable scope of our attack.

\subsection{Transferability of Our Attack}
\label{subsection: transferability of our attack}
In this section, we consider the transferability of our attack, i.e., the attack performance when testing negatives of the attack model are of different datasets from training negatives, and evaluate it on conditional generative tasks.
Concretely, the attack model is established following the settings in \cref{subsection: evaluation settings}, i.e., to construct an auxiliary dataset according to the member one, and then to generate/sample training positives/negatives.
However, during the testing time, we do not sample negative images from the auxiliary dataset, but from another non-member dataset, i.e., LSUN-Bedroom in this case.
As shown in \cref{subfigure: conditional results} in red bars, the transferability performance is comparable to the attack performance in \cref{subsection: results}, and obviously superior to the black-box baseline~\cite{CYZF20}.
Even requiring fewer assumptions than the white-box baseline, in most cases, our attack can achieve stronger transferability performance.
For instance, the transferability performance of VQGAN is an AUC of 0.956 while the black- and white-box baselines only achieve an AUC of 0.554 and 0.583, respectively.
These significant advantages further indicate the effectiveness of our attack.

\subsection{Influence of the Query Budget}
\label{subsection: influence of the query budget}

\begin{figure*}[!t]
\centering
\subfloat[(CIFAR-10, STL-10).]{
    \centering
    \label{subfigure: the number of generated cifar10}
    \includegraphics[width=0.45\textwidth]{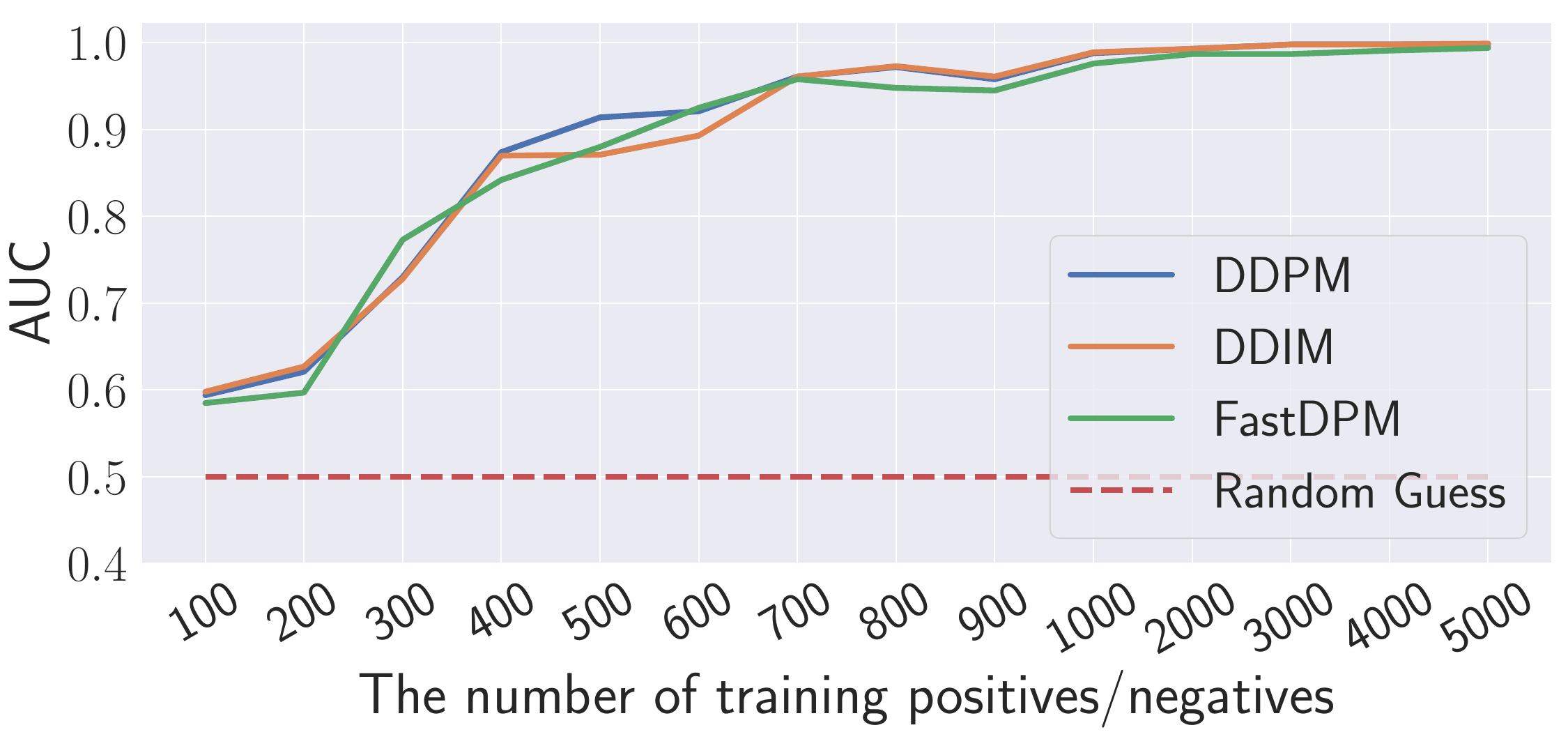}
}
\subfloat[(CelebA, UTKFace).]{
    \centering
    \label{subfigure: the number of generated celeba}
    \includegraphics[width=0.45\textwidth]{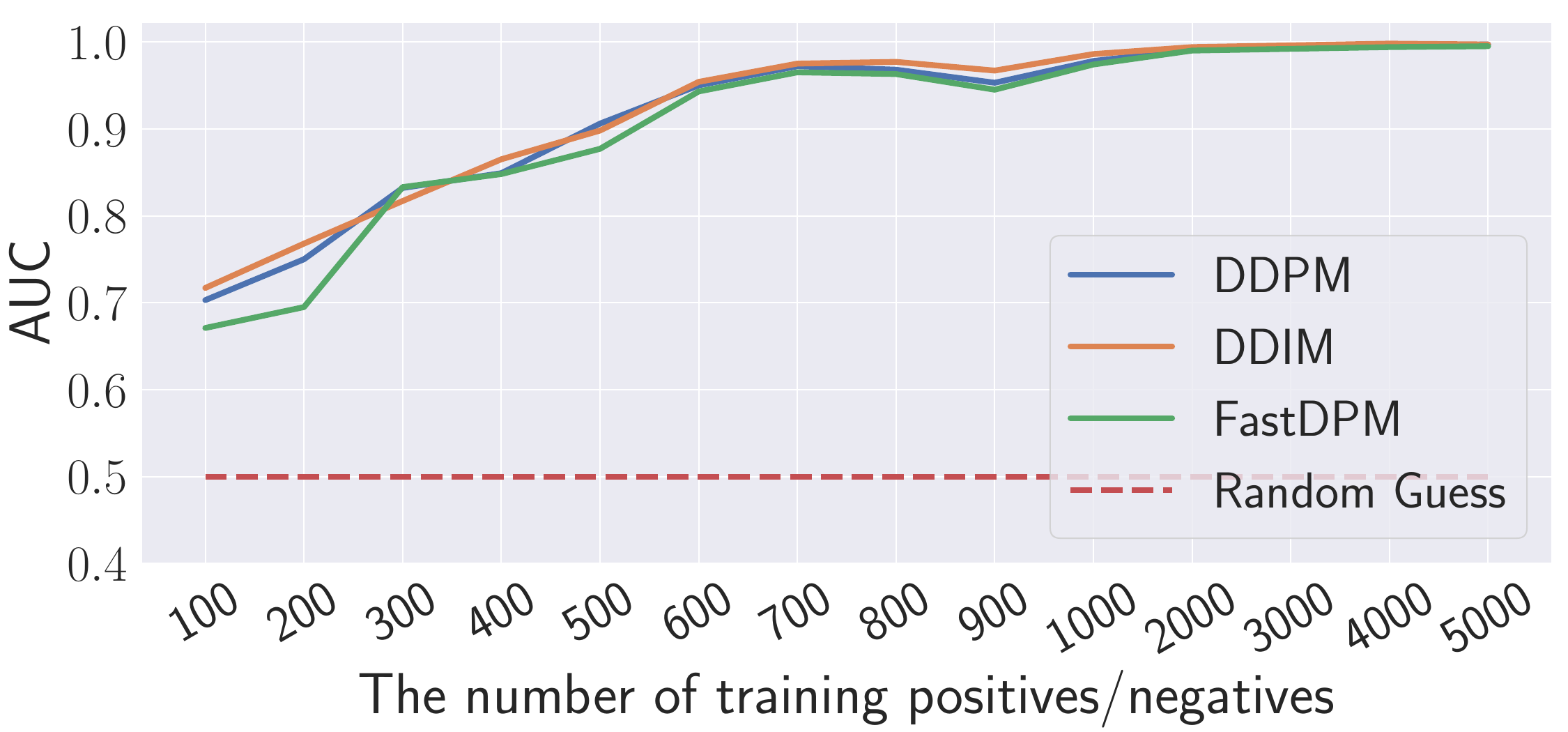}
}
\caption{The attack performances with different query budgets in unconditional generation tasks.}
\label{figure: the number of generated samples}
\end{figure*}

We further evaluate the influence of the query budget on the attack performance.
Sometimes users will be required to pay for the query, limiting the adversary's ability to generate unlimited images (which also influences the domains of auxiliary datasets as mentioned in \cref{subsection: auxiliary datasets}).
And the generation cost is especially expensive for DDPMs compared to other generative models.
Thus in this section, we mainly focus on how the query budget influences the attack performance on unconditional generation tasks implemented by DDPMs.
Specifically, we change the number of generated images (i.e., training positives), from 100 to 1,000 with a step size of 100 and from 1,000 to 5,000 with a step size of 1,000, and see the impact on the attack performance.
Note that, to balance the training positives and negatives, we also control for an equal number of images sampled from the auxiliary dataset.
As \cref{figure: the number of generated samples} shows, the attack performance advances when the number of generated images increases.
Then the attack performance saturates when the number of training positives/negatives arrives at 700.
Interestingly, even though the generation quantity is only 100, our attack is still effective in most cases.
For instance, when targeting DDPM trained on CelebA, our attack achieves an AUC of 0.703 while the black-box baseline only achieves 0.527, as shown in \cref{subfigure: the number of generated celeba}.
This observation indicates our attack is applicable even when the query budget is insufficient.

%------------------------------------------------------------------------
\section{Discussion}
\label{section: discussion}

\begin{figure*}[!t]
\centering
\subfloat[Unconditional generation tasks.]{
    \centering
    \label{subfigure: more nonmembers unconditional}
    \includegraphics[width=0.45\textwidth]{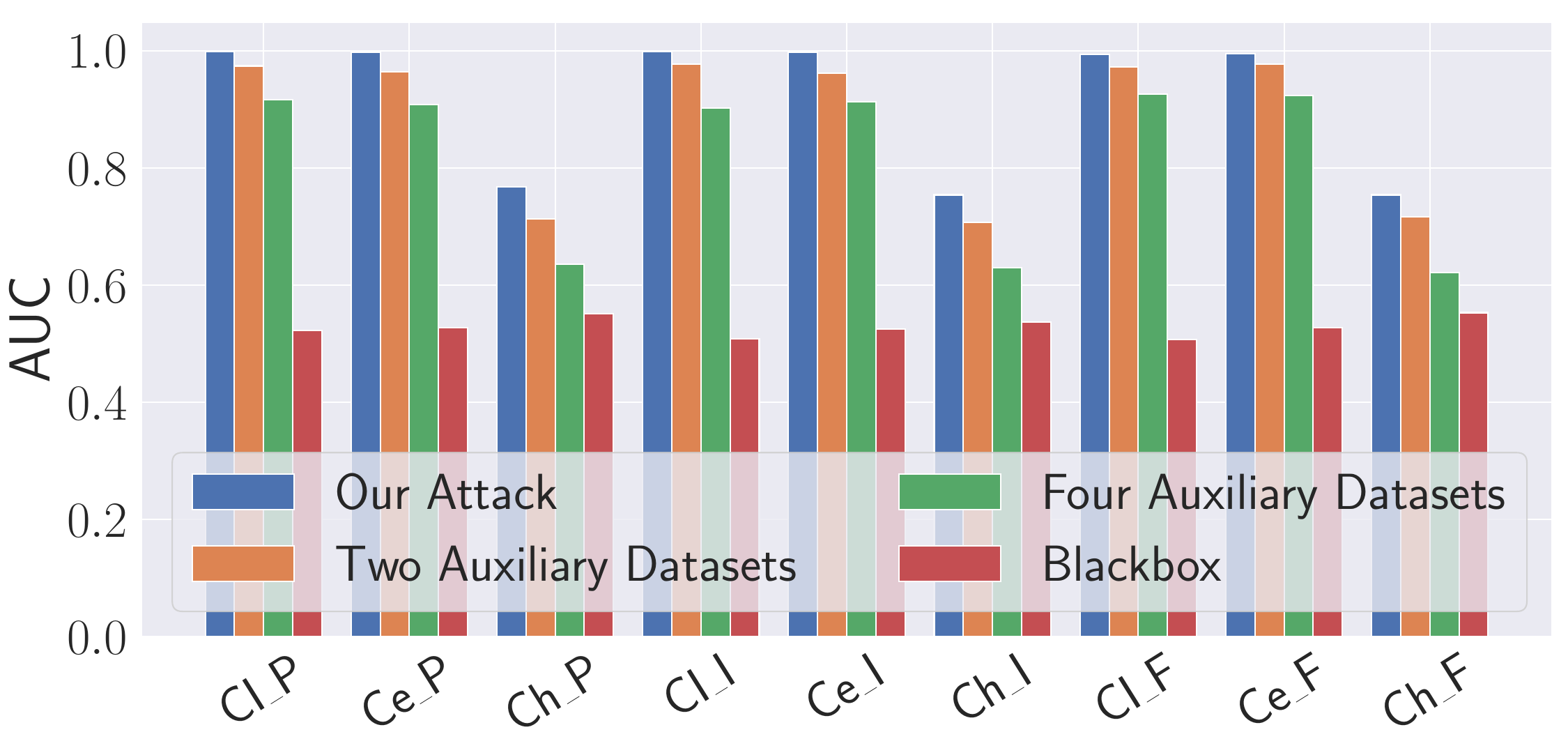}
}
\subfloat[Conditional generation tasks.]{
    \centering
    \label{subfigure: more nonmembers conditional}
    \includegraphics[width=0.45\textwidth]{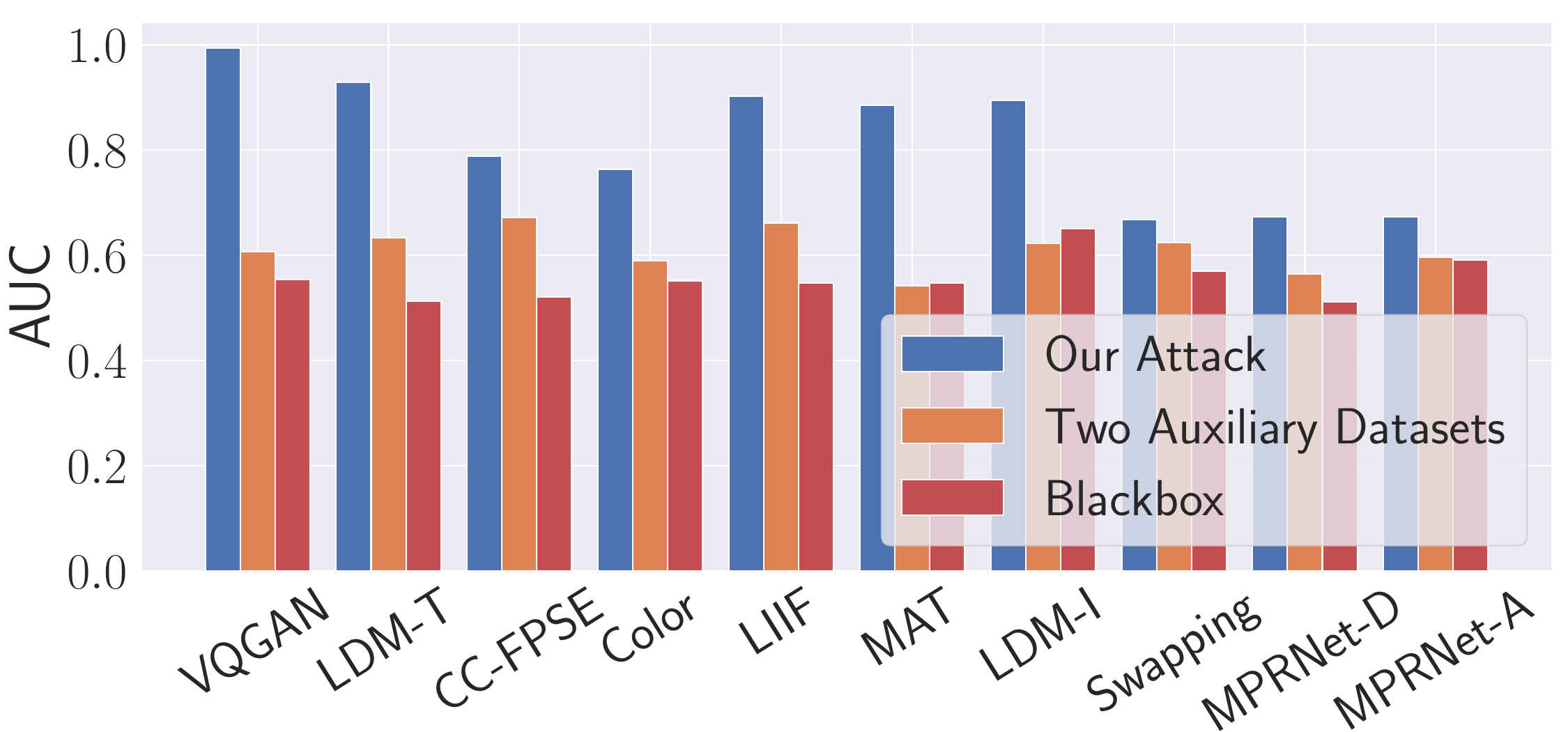}
}
\caption{The attack performances with different numbers of auxiliary datasets, where ``Two Auxiliary Datasets'' represents the original auxiliary dataset described in \cref{table: external scenarios} and Wild-Bedroom.
Besides, ``Four Auxiliary Datasets'' in \cref{subfigure: more nonmembers unconditional} represents STL-10, UTKFace, Wild-Bedroom, and Wild-Church.
Regarding the x-axis, the setting is the same as \cref{figure: effective on generative applications}.}
\label{figure: more nonmembers}
\end{figure*}

\mypara{Membership Boundary}
As mentioned in \cref{subsection: auxiliary datasets}, all samples that are not used to train the target generator are non-members.
That means the boundary between members and non-members actually separates members and all other samples.
However, in \cref{section: evaluation}, we only consider a single auxiliary dataset as non-members (i.e., training and testing negatives).
Thus, to better understand our work, we evaluate cases where more auxiliary datasets are involved as non-members.
As shown in \cref{figure: more nonmembers}, with more auxiliary datasets, our attack performance drops but is still strong (especially in unconditional generation tasks as depicted in \cref{subfigure: more nonmembers unconditional}) and significantly better than the black-box baseline~\cite{CYZF20} in most cases.
This decrease is because more auxiliary datasets (i.e., non-members) will make the boundary more complicated and then harder to learn by the attack model.
This finding leaves an interesting future work on dealing with complicated boundaries.

\begin{figure}[!t]
\centering
\includegraphics[width=0.45\textwidth]{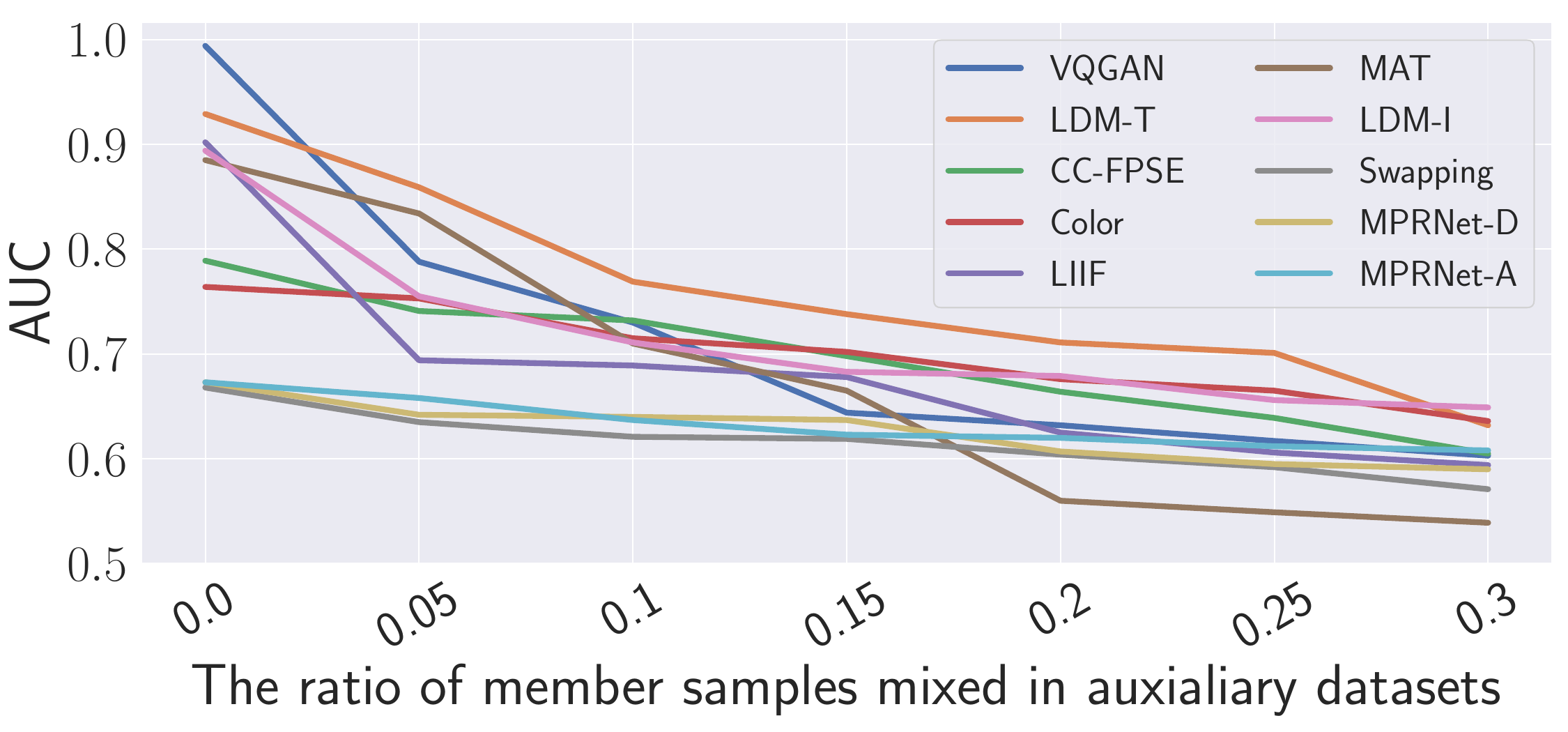}
\caption{The attack performances with different ratios of overlap between members and auxiliary datasets.}
\label{figure: overlap mix}
\end{figure}

\mypara{Overlap Between Auxiliary and Member Samples}
As mentioned in \cref{subsection: auxiliary datasets}, ideally, there is no overlap between auxiliary and corresponding member samples.
However, since members are inaccessible, non-overlap is hard to guarantee in practice.
Thus, we here explore the influence of overlaps on our attack by mixing member samples into auxiliary datasets by different ratios.
We can see from \cref{figure: overlap mix} that as expected, the smaller the overlap, the better the attack performance.
Surprisingly, even when a 0.1 ratio of members is mixed into auxiliary datasets, our attack can still achieve an AUC $> 0.7$ in most cases, which further indicates the practicability and effectiveness of our work.

\begin{table}[!t]
    \caption{The attack performances with different sources of training negatives, i.e., sampled from the auxiliary dataset or generated by a shadow model.}
    \resizebox{0.47\textwidth}{!}{
    \centering
    \begin{tabular}{l | c c c c | c c}
        \toprule
        \multirow{2}{*}{\textbf{Application}} & \multicolumn{2}{c}{\textbf{Technique}} & \multicolumn{2}{c|}{\textbf{Dataset}} & \multicolumn{2}{c}{\textbf{Negatives}} \\
        & \textbf{Target} & \textbf{Shadow} & \textbf{Member} & \textbf{Auxiliary} & \textbf{Sampled} & \textbf{Generated} \\
        \midrule
        Text-conditional & \multirow{2}{*}{LDM} & \multirow{2}{*}{CC-FPSE} & \multirow{2}{*}{LAION} & \multirow{2}{*}{COCO} & 0.929 & \underline{0.938} \\
        Image inpainting & & & & & \underline{0.894} & 0.741 \\
        \midrule
        \multirow{2}{*}{Unconditional} & DDPM & \multirow{2}{*}{FastDPM} & \multirow{2}{*}{CelebA} & \multirow{2}{*}{UTKFace} & \underline{0.997} & 0.995 \\
        & DDIM & & & & \underline{0.997} & 0.996 \\
        \bottomrule
    \end{tabular}}
    \label{table: training negative}
\end{table}

\mypara{Generation of Training Negatives}
As described in \cref{subsection: methodology}, training negatives are sampled from auxiliary datasets while training positives are generated by generative models.
Thus, we wonder about the feasibility of generating the training negatives for our attack model by a different generative model (i.e., a \emph{shadow model}) from the target generator, which also provides a guarantee of non-overlapping, e.g., training negatives of COCO (UTKFace) are generated by CC-FPSE (FastDPM).
From \cref{table: training negative} we can see that generating training negatives achieves a comparable performance to our attack, yet an extra cost of generation will be required.
This indicates the practicability of simply sampling training negatives from auxiliary datasets, unless aiming at the non-overlapping guarantee.
On the other hand, the attack performance is shown sensitive to the choice of auxiliary datasets (see the first part in \cref{section: discussion}), however, finding suitable auxiliary samples in practice might be challenging even though we only require a similar distribution instead of the same one.
To this end, involving shadow models to generate training negatives would help this problem.

\begin{table}[!t]
    \caption{The attack performances regarding 
    of members and non-members being from the same distribution, and measuring by the metric proposed by \cite{CCNSTT21}.}
    \centering
    \resizebox{0.47\textwidth}{!}{
    \centering
    \begin{tabular}{l | c | c c | c c c c c c}
        \toprule
        \multirow{3}{*}{\textbf{Technique}} & \multirow{3}{*}{\textbf{Dataset}} & \multicolumn{2}{c|}{\textbf{Same Distribution}} & \multicolumn{6}{c}{\textbf{Evaluation Metric}} \\
        & & \multirow{2}{*}{\textbf{Ours}} & \multirow{2}{*}{\textbf{SecMI}} & \multicolumn{2}{c}{\textbf{Ours}} & \multicolumn{2}{c}{\textbf{Blackbox}} & \multicolumn{2}{c}{\textbf{Whitebox}} \\
        & & & & \textbf{TPR} & \textbf{FPR} & \textbf{TPR} & \textbf{FPR} & \textbf{TPR} & \textbf{FPR} \\
        \midrule
        DDPM & \multirow{2}{*}{CelebA} & \textbf{0.545} & 0.516 & \textbf{0.998} & \textbf{0.002} & 0.517 & 0.472 & 0.606 & 0.397 \\
        DDIM & & \textbf{0.543} & 0.519 & \textbf{0.997} & \textbf{0.002} & 0.515 & 0.480 & 0.604 & 0.402 \\
        \bottomrule
    \end{tabular}}
    \label{table: definition and evaluation}
\end{table}

\mypara{Membership Status}
To further empirically support our method, we now evaluate a stricter assumption where members and non-members are disjointly from the same dataset.
Specifically, we use the training and testing samples of CelebA as members and non-members on DDPM and DDIM respectively as shown in the “Same Distribution” column of \cref{table: definition and evaluation}.
Compared to a recent MIA against diffusion models, i.e., SecMI~\cite{SecMI}, our method shows obviously better attack performance.
This result suggests a more comprehensive application scenario of our method.

\mypara{Evaluation Metric}
Carlini et al. propose that the correctness of inferring members is more practically meaningful than non-members~\cite{CCNSTT21}.
Thus, we evaluate our attack by true positive rate (TPR) and false positive rate (FPR), where a higher TPR and a lower FPR derive a better membership inference.
In the ``Evaluation Metric'' column of \cref{table: definition and evaluation}, we can see that our method could gain an obviously higher TPR and lower FPR even though the accuracy gap between ours and Whitebox is not as significant (\cref{subfigure: unconditional results}), indicating the multi-faceted effectiveness of our method.

%------------------------------------------------------------------------
\section{Conclusion}
\label{section: conclusion}

Generative models increasingly show their promising talents in generating realistic and creative images.
However, the privacy risks of training data leakage introduced by them are largely unexplored.
Previous membership inference attacks have proved that generative models are vulnerable to privacy leakage by inferring whether a query sample is in the training dataset.
However, the existing works require shadow models and/or white-box access, and ignore or only focus on the state-of-the-art DDPMs, which limits their application scope.
In this paper, we propose the first generalized membership inference attack against various generative models including GANs, [V]AEs, IFs, and DDPMs.
We only assume the adversary can obtain generated distributions from target generators.
Under the black-box setting, our attack is agnostic to the architectures and applications of generative models.
Extensive experimental results show the effectiveness of our attack against various generative models and applications.
Further studies show that our attack still works with a limited query budget, and the transferability makes our attack a real threat in real-world scenarios.
Consequently, we aim to call for community awareness of the privacy protection of generative models.

%------------------------------------------------------------------------
{\small
\bibliographystyle{ieee_fullname}
\bibliography{normal_mybib}
}

%------------------------------------------------------------------------
\appendix

\section{Datasets}
\label{appendix: datasets}

In this section, we will introduce the datasets used in our experiments, including CIFAR-10, STL-10, CelebA, UTKFace, LSUN, ImageNet, Open Images, LAION, COCO, ADE20K, and SIDD.

\begin{itemize}
    \item \textbf{CIFAR-10.}
    The CIFAR-10 dataset consists of $60,000$ images of 10 classes, i.e., airplane, automobile, bird, cat, deer, dog, frog, horse, ship, and truck.
    For each class, there are $5,000$ training images and $1,000$ test images.
    \item \textbf{STL-10.}
    The STL-10 dataset is inspired by the CIFAR-10 dataset but with some modifications, covering the classes of airplane, bird, car, cat, deer, dog, horse, monkey, ship, and truck.
    And for each class, there are $500$ training images and $800$ test images.
    Besides, there are $100,000$ unlabeled images for unsupervised learning.
    \item \textbf{CelebA.}
    The CelebA dataset contains more than $200,000$ celebrity images with different face attributes.
    Each face image has 40 attribute annotations with binary values, indicating whether this image satisfies corresponding attributes, e.g., whether wearing a hat.
    \item \textbf{UTKFace.}
    The UTKFace dataset contains $20,000$ face images with annotations of age, gender, and ethnicity, with a long age span (range from 0 to 116 years old).
    \item \textbf{LSUN.}
    The LSUN dataset contains 10 scene categories, e.g., bedroom and church.
    For LSUN-Bedroom, there are more than $3,000,000$ training images and $300$ validation images.
    And for LSUN-Church, there are more than $120,000$ training images and $300$ validation images.
    \item \textbf{ImageNet.}
    The ImageNet dataset contains more than $14,000,000$ annotated images according to the WorldNet hierarchy.
    In the paper, we use its tiny version, i.e., Tiny ImageNet.
    The Tiny ImageNet dataset covers 200 classes, and each class has $500$ training images, $50$ test images, and $50$ validation images.
    \item \textbf{Open Images.}
    The Open Images dataset consists of 9 million training images, which are partially annotated, with $9,600$ trainable classes.
    \item \textbf{LAION.}
    The LAION-400M dataset contains 400 million English (image, text) pairs.
    And all images and texts are filtered by CLIP~\cite{RKHRGASAMCKS21} according to the cosine similarity between the image and text embeddings.
    \item \textbf{COCO.}
    The COCO dataset is a large-scale dataset containing $330,000$ images, within which more than $200,000$ images are labeled with 80 object categories.
    \item \textbf{ADE20K.}
    The ADE20K dataset contains more than $20,000$ scene-centric images which are annotated with pixel-level objects and object parts labels.
    There are totally 150 semantic categories, including stuffs and discrete objects.
    \item \textbf{SIDD.}
    The SIDD dataset consists of $30,000$ noisy images taken by 5 representative smartphone cameras from 10 scenes under different lighting conditions.
\end{itemize}

%------------------------------------------------------------------------
\section{Training Details}
\label{appendix: training details}

For label balance, the number of training positives is equal to that of training negative in our experiments.
If there is no limitation on query times, it is possible to generate any desired number of training positives.
Thus, the number of training samples relies on the size of auxiliary datasets (where the training negatives are derived).
In our experiments, in any case that the auxiliary dataset could provide more than $5,000$ training negatives, we will set the number of training samples to $10,000$ (i.e., $5,000$ for training positives and $5,000$ for training negatives).

%------------------------------------------------------------------------
\end{document}